\documentclass[aps,pre, twocolumn,float,amsmath]{revtex4}
\usepackage[dvips]{color}
\usepackage[normalem]{ulem} 
\definecolor{g-blue}{rgb}{0.83,0.95,1}
\definecolor{g-yellow}{rgb}{1,1,0.7}
\definecolor{g-green}{rgb}{0.9,1,0.9}
\definecolor{green}{rgb}{0,0.6,0}
\definecolor{cyan}{rgb}{0,0.7,0.7}
\definecolor{black}{rgb}{0,0,0}
\definecolor{grey}{rgb}{0.4 ,0.4 ,0.4 }

\usepackage{bm}
\usepackage{amssymb}
\usepackage{amsfonts}
\usepackage{amsmath}
 \usepackage{graphicx}
  \usepackage{amsmath,bm,epsfig}

\newcommand{\Eq}[1]{Eq.\,(\ref{#1})}
\newcommand{\Fig}[1]{Fig.\,\ref{#1}}
\newcommand{\Ref}[1]{Ref.\,\cite{#1}}

\def\be{\begin{equation}}\def\ee{\end{equation}}
\def\bea{\begin{eqnarray}}\def\eea{\end{eqnarray}}
\def\bse{\begin{subequations}}\def\ese{\end{subequations}}
\newcommand{\BE}[1]{\begin{equation}\label{#1}}
\newcommand{\BEA}[1]{\begin{eqnarray}\label{#1}}
\newcommand{\BSE}[1]{\begin{subequations}\label{#1}}

\let \nn  \nonumber

\let \= \equiv \let\*\cdot \let\~\widetilde \let\^\widehat \let\-\overline

  \def\1{\bm1} 

\def\<{\left\langle}    \def\>{\right\rangle}
\def\({\left(}          \def\){\right)}
 \def \[ {\left [} \def \] {\right ]}

\renewcommand{\a}{\alpha}

\newcommand{\B}[1]{{\bm{#1}}}
\newcommand{\C}[1]{{\mathcal{#1}}}    

\renewcommand{\sb}[1]{_{\text {#1}}}  
\renewcommand{\sp}[1]{^{\text {#1}}}  

\def\He4 {$^4$He~}

\begin{document}
\title{Reply to Comment on ``Dynamics of the Density of Quantized Vortex-Lines in Superfluid Turbulence"  }
\author{D. Khomenko,  V. S. L'vov,  A. Pomyalov,  and I. Procaccia}
\affiliation{Department of Chemical Physics,  Weizmann Institute  of Science, Rehovot 76100, Israel }
\begin{abstract}
This is a Reply to Nemirovskii Comment [Phys. Rev. B 94, 146501 (2016)] on the Khomenko \emph{et al}[Phys.Rev. B {\bf 91}, 180504(2016)], in which a new form of the production term  in  Vinen's equation for the evolution of the vortex-line density $\C L$ in the thermal counterflow  of superfluid $^4$He  in a channel  was suggested.  To further substantiate the suggested form which was questioned in the Comment, we present a physical explanation for the improvement of the  closure  suggested in Khomenko \emph{et al}[Phys.Rev. B {\bf 91}, 180504(2016)] in comparison to the form proposed by Vinen. We also discuss the  closure  for the flux term, which  agrees  well  with the  numerical results without any fitting parameters.

\end{abstract}
\maketitle

A complete dynamical theory of quantized vortex tangles continues to be a challenge, requiring careful attention to numerous details \cite{SN,PRB,4,5,6}. The Comment by S. Nemirovskii (hereafter refereed to as  SN) requires an additional clarification of the approximations and assumptions made in our approach\,\cite{PRB} to the dynamics of the vortex line density $\C L(\B r, t)$ in superfluid turbulence.

The vorticity in superfluid $^4$He is quantized: It is constrained to vortex-line singularities of fixed circulation $\kappa=h/M$, where $h$ is Planck's constant and $M$ is the mass of the $^4$He atom. The smallness of the  vortex-line core radius
$a_0\simeq 10^{-8}\,$cm allows us to consider them  as directional geometrical lines $\B s(\xi, t)$, traditionally parametrized\,\cite{4} by an arclength   $\xi$.

A good {\em phenomenological} starting point in the analysis of the vortex line dynamics  is the Vinen equation\,\cite{1} for the evolution of $\C L(t)$ in space homogeneous (thermally driven) superfluid $^4$He flow with a counterflow velocity $V\sb{ns}$,
\begin{equation}\label{VE}
\frac{d \C L(t)}{dt}= \C P(t)- \C D(t)\ .
\end{equation}
Here the production and the decay terms, $\C P(t)$ and  $\C D(t)$  are expressed in terms of $\C L(t)$ and $V\sb {ns}$ only.

\begin{figure}
   \includegraphics[scale=0.4 ]{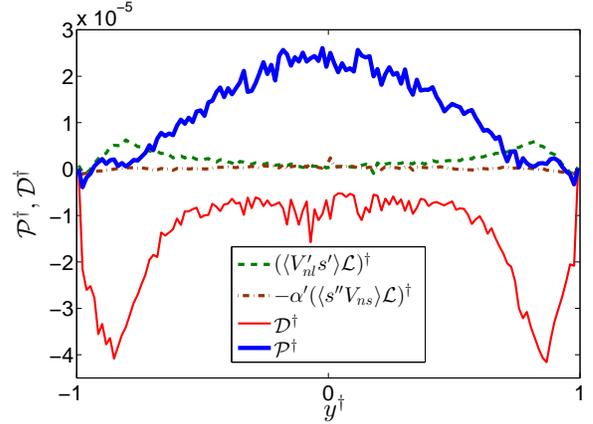}
\caption{  \label{f:1}  Color Online. Relative contributions of the neglected terms in the microscopic vortex dynamics \eqref{gen} to $\partial \C L(\B r,t)/\partial t $  are shown by dashed and dot-dashed lines. The accounted for contributions to $\C P$ and $\C D$ are shown by the thick and thin solid lines, respectively.  All quantities are normalized according to Eq. \eqref{norm}. (Parabolic profile, $T=1.6\,$K).  }
\end{figure}

In \Ref{PRB} we reconsidered the evolution equation for $\partial \C L(\B r,t)/\partial t $ starting from the {\em microscopic} Schwarz equation \cite{4,5} for the length of the vortex-line segment $\delta \xi$. This equation contains two temperature dependent dimensionless mutual friction parameters $\a$ and $\a'$,
\begin{equation}\label{gen}
\frac{1}{\delta \xi}\frac{d\delta \xi }{dt} =  \alpha V\sb{ns}(\B s,t)\cdot (\B s'\times \B s'')+\B s'\cdot {\B V\sb{nl}^s}'-\alpha' \B s''\cdot \B V\sb{ns} \ .
\end{equation}
 We agree with SN that this equation includes more terms than  {Eq.(4a) of \Ref{PRB}}.
For that reason we used in  Eq.(4a) an ``$\approx$" sign instead of an ``$=$" to indicate  that only the most dominant  contributions were retained for the case of counterflow turbulence in a channel. The relative importance of the different terms of \eqref{gen} is shown in \Fig{f:1}  using the following normalization:
 \begin{equation}\label{norm}
y^{\dagger}=y/H, \C P^{\dagger}=\kappa^3\C P/\langle V\sb{ns}^2\rangle^2, \C D^{\dagger}=\kappa^2\C J/\langle V\sb{ns}^2\rangle^{3/2}.
\end{equation}

\begin{figure}
    \includegraphics[scale=0.4]{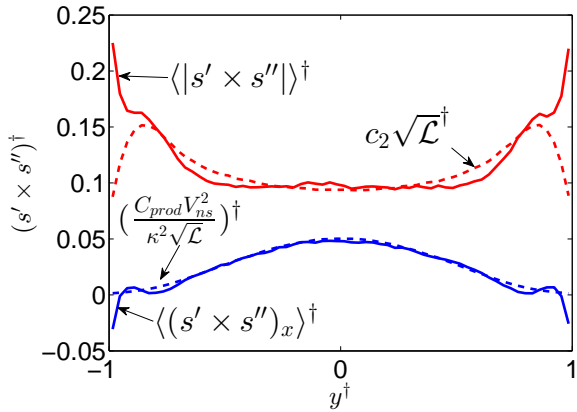}
\caption{  \label{f:2}  Color Online. Comparison of the streamline projection of $\< \B s' \times \B s''\> $ (blue line) with its modulus (red line) for the parabolic profile. The blue and red dashed lines denote properly normalized profiles of $V\sb{ns}^2/ \sqrt {\C L}$ and $\sqrt {\C L}$.  All quantities are normalized by $\sqrt{\<V_{ns}^2\>}/\kappa$.}
\end{figure}

  We agree with SN that any decomposition of superfluid velocities may be considered arbitrary. We decided to  follow Schwarz\,\cite{5} and to  assign the contribution proportional to $|s''|^2$ to the decay term. There are two reasons for that. First, this term should remain active after switching off the counterflow and thus should not vanish when $V\sb{ns}=0$.  Second, it should be positive definite. All the other contributions were included in the production term.
Having clarified  {the form of  \Eq{gen} in \Ref{PRB}}, we get straightforwardly to  Eq\,(6a) in \Ref{PRB} for the production term:
\begin{subequations} \label{prod}
\begin{equation} \label{prodA}
 \C P (y,t)= \alpha  \C L(y)   \<
  \B V\sb{ns}
  \cdot (\B s'\times\B s'')\>_{x,z}\ .
 \end{equation}
 Here $\< \cdot \cdot \cdot  \>_{x,z}$ denotes an $y$-dependent average over the vortex tangle residing in thin slices of width $\delta y$ parallel to the channel wall.
 \Eq{prodA} involves a smooth macroscopic field $\B V\sb {ns}$  that changes slowly along the vortex line and a factor $(\B s'\times\B s'')$ that is defined by a local structure of the tangle and changes fast along the vortex line. That allows us to take the slow [\,$(y,t)$-dependent\,] factor $\B V\sb {ns} $ out of the average and account for the scalar product by retaining only a streamwise projection of $(\B s'\times\B s'')$:
\begin{eqnarray}\nn \label{prod2}
 \C P (y,t)&\approx& \alpha  \C L(y)
  \B V\sb{ns}(y) \cdot\<  \B s'\times\B s'' \>_{x,z}\\
&=& \alpha  \C L(y)  V\sb{ns}(y) \< (\B s'\times\B s'')_x\>_{x,z} \\\nonumber
\end{eqnarray}
 \end{subequations}
 We see that rigorous analysis of the production term\,\eqref{prod2} in  Eq.\eqref{VE} for $d \C L(t)/ dt$ requires an equation for the streamwise projection $ (\B s'\times\B s'')_x$.
   One can find it  analyzing an equation for  $ d \< (\B s'\times\B s'')_x\>_{x,z}/ dt $, which can be derived from \Eq{gen}. Unfortunately, the equation for $ d \< (\B s'\times\B s'')_x\>_{x,z}/ dt $   involves even more complicated statistical characteristics of the random vortex tangle. A way to close such an infinite \emph{unclosed} chain of coupled equations is to find a proper \emph{closure} approximation (hereafter referred to for shortness as \emph{closure}) which expresses $ (\B s'\times\B s'')_x$ in terms of macroscopic objects $V\sb{ns}$ and $\C L$.

   A possible closure for the streamwise  projection $ (\B s'\times\B s'')_x$ follows from the observation that its
  value is equal to that of the local curvature: $|\B s'\times \B s''|= |\B s''|$.  Assuming for a moment  that on  average
  \begin{equation}\label{as1}
  \<(\B s'\times\B s'')_x\>_{x,z} \simeq \< |s''|\>_{x,z}\,,
   \end{equation}
   and accepting the common and reasonably well-justified  approximation that $\< |s''|\>_{x,z}\propto \sqrt{\C L}$ (see the solid and dashed red lines in \Fig{f:2}), we end up with  Vinen's form of the  production term,
    \begin{subequations}\label{prod1}
    \begin{equation}\label{prod-1}
    \C P_1\sim \alpha V\sb{ns} \C L^{3/2}\ .
    \end{equation}
Note however that the assumption\,\eqref{as1} is not fulfilled even on a qualitative level. On one hand, from the theoretical view point, Eq.\,\eqref{as1} is in contradiction with simple symmetry considerations: In the absence of the counterflow there is no preferred direction in the problem (far away from the wall) and one expects $\<(\B s'\times\B s'')_x\>_{x,z}=0$, whereas  $\< |s''|\>_{x,z}$ has a well defined value.
On the other hand, comparing  in \Fig{f:2} the calculated wall-normal profiles of  $\<(\B s'\times\B s'')_x\>_{x,z}$ (blue line) and $\<| \B s'\times\B s''|\>_{x,z}$ (red line), we find that these profiles demonstrate completely different behavior.

\begin{figure*}
\begin{tabular}{cc}
(a)& (b)\\
   \includegraphics[scale=0.4 ]{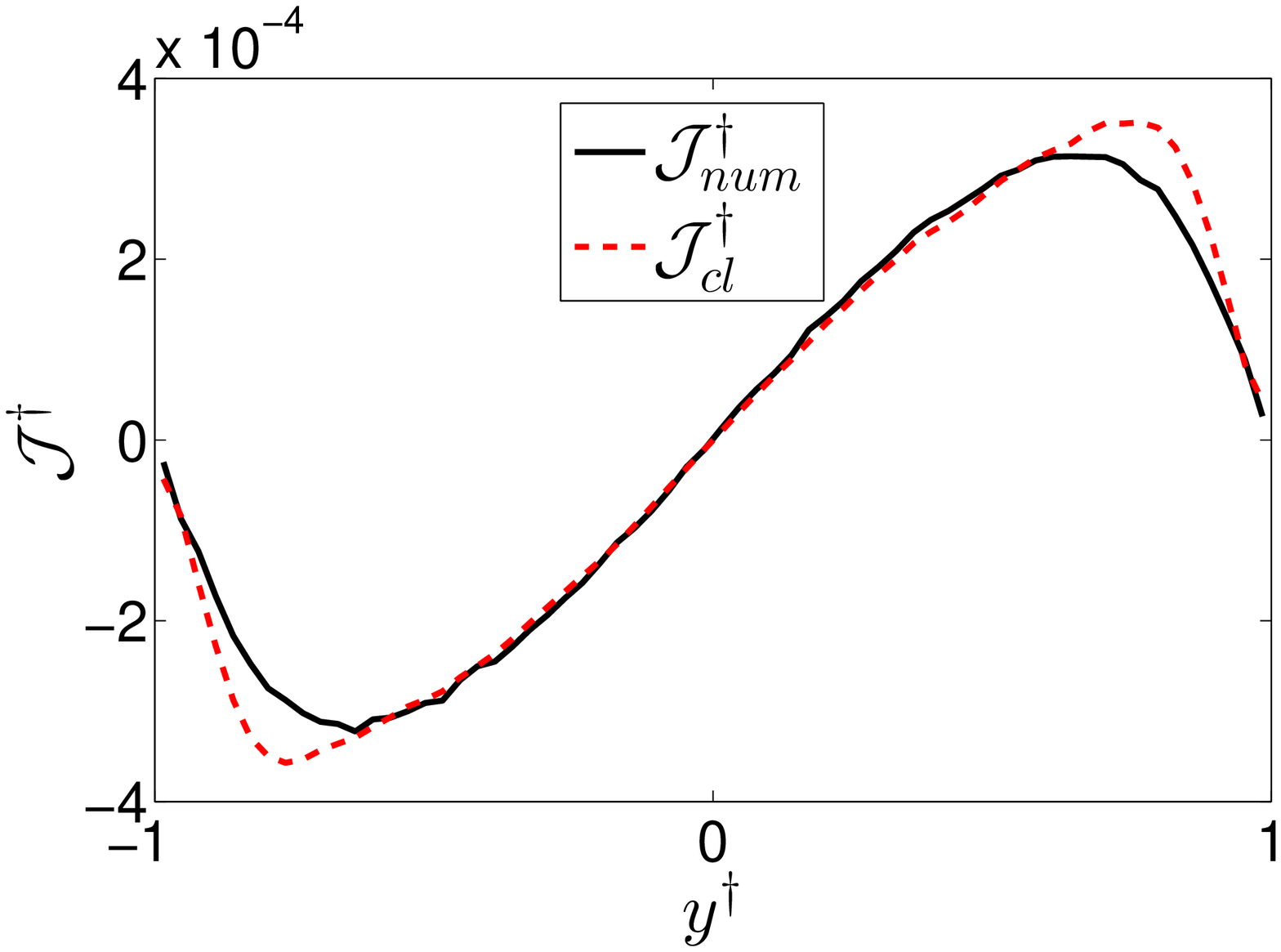}&
 \includegraphics[scale=0.4 ]{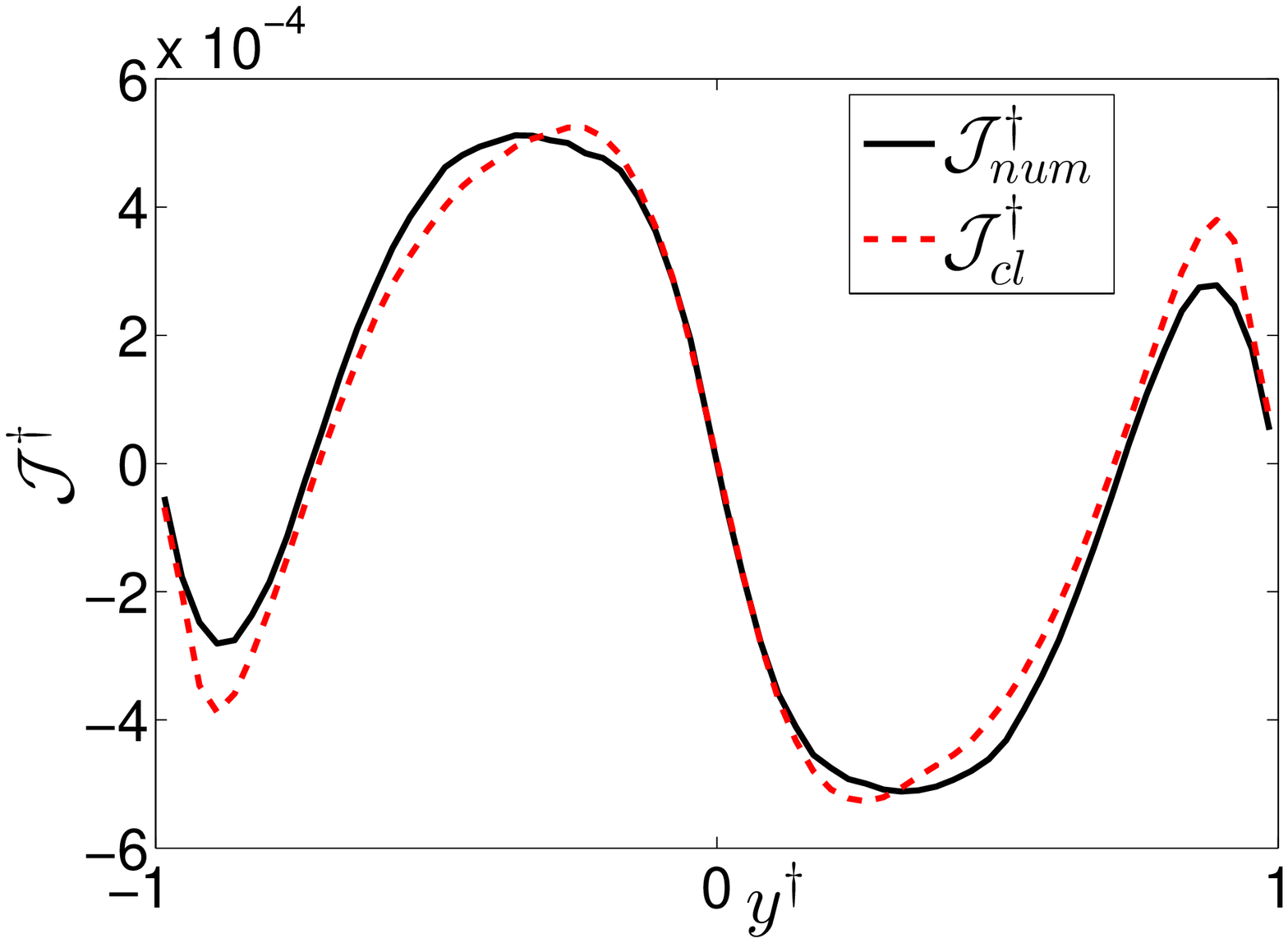}\\
\end{tabular}
\caption{  \label{f:4}  Color Online. Comparison of numerically measured flux and its modeling. Panel a: parabolic profile. Panel b: non-parabolic profile. The solid lines denote numerical results, and the dashed lines correspond to the  closure \eqref{flux}.  All quantities are normalized according to Eq.\eqref{norm}. }
\end{figure*}

\begin{figure*}
\begin{tabular}{cc}
(a)& (b)\\
   \includegraphics[scale=0.4 ]{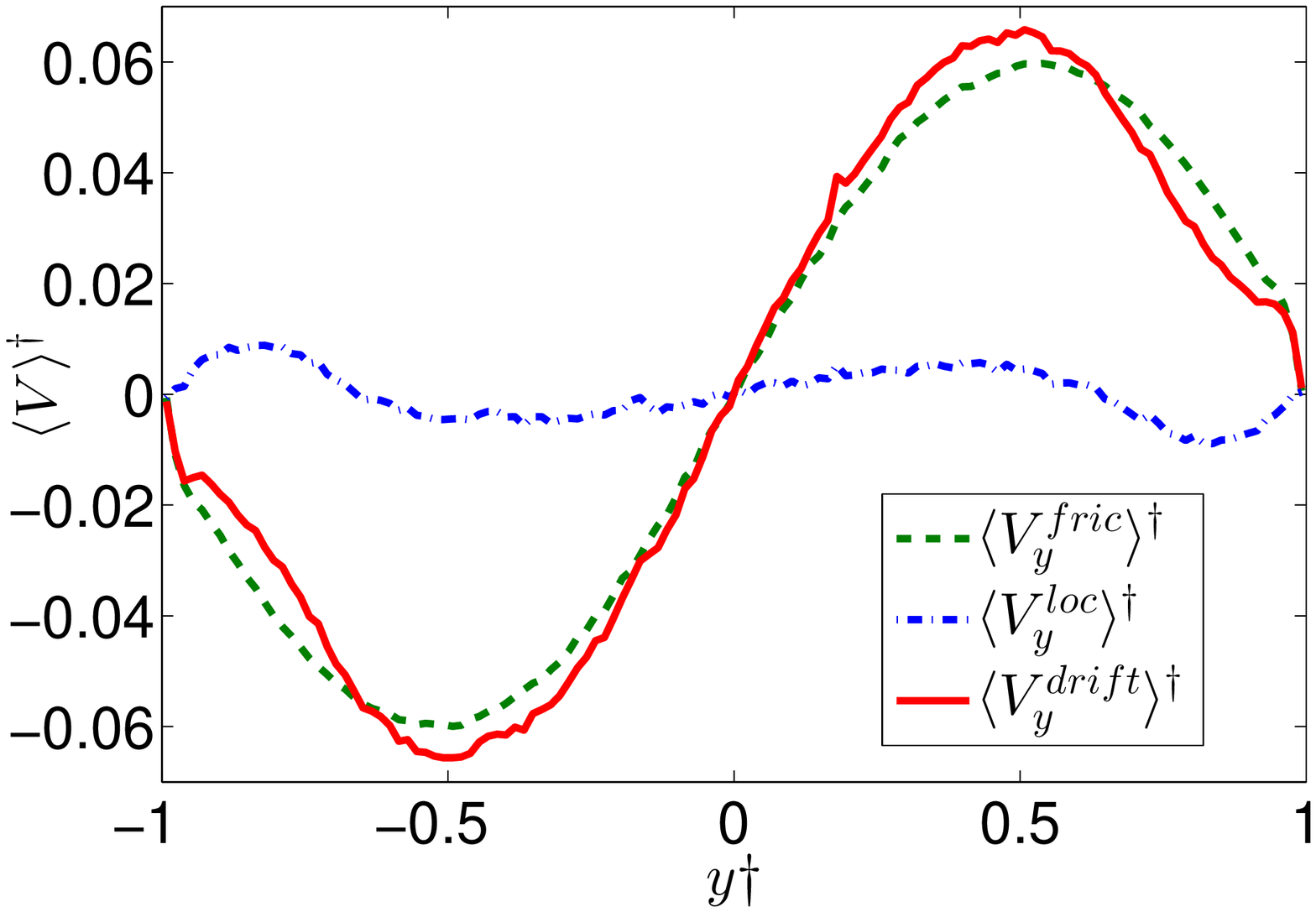}&
 \includegraphics[scale=0.4 ]{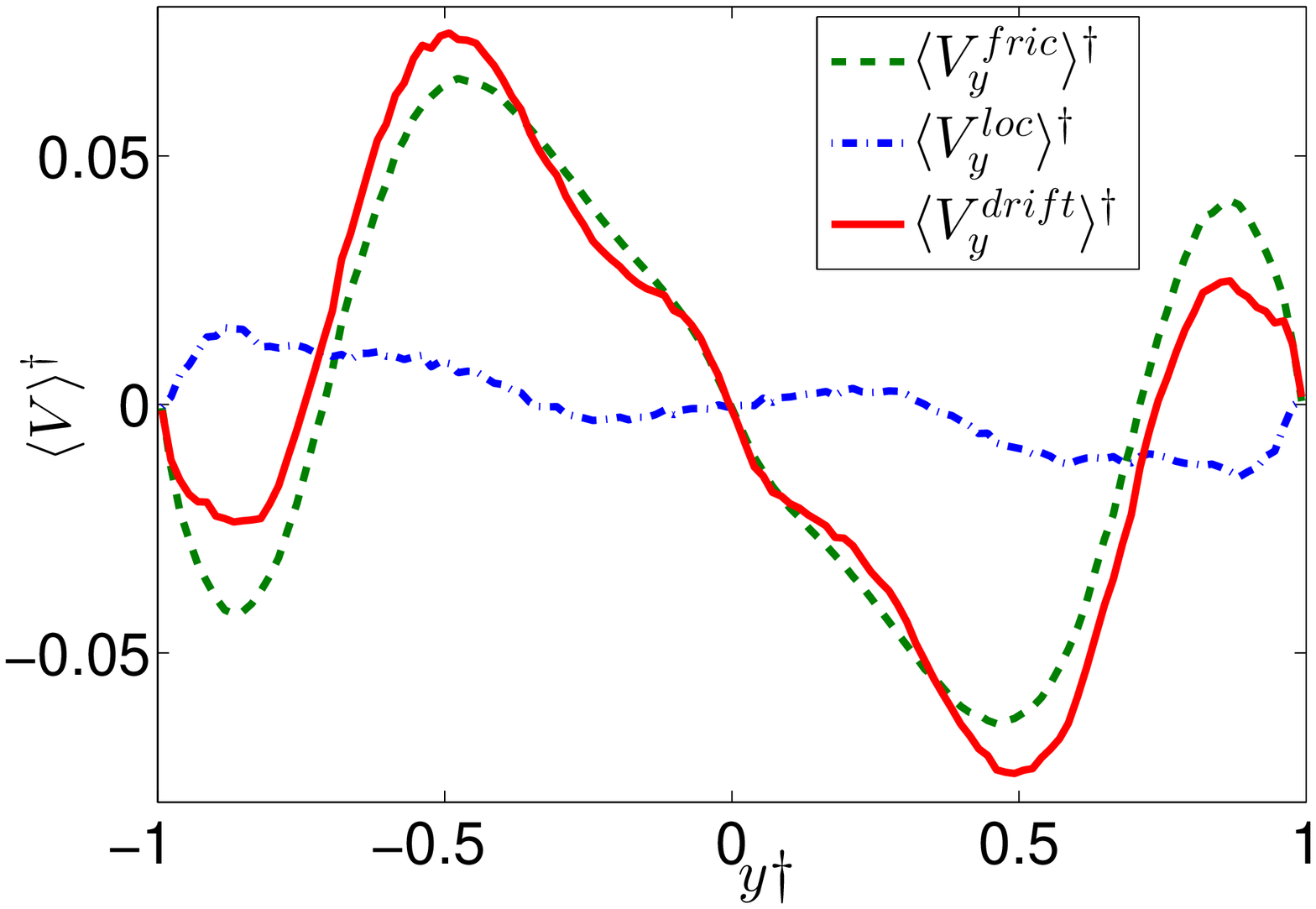}\\
\end{tabular}
\caption{  \label{f:3}  Color Online. Comparison of the components of the drift velocity. Panel a: parabolic profile. Panel b: non-parabolic profile. The velocities are normalized by $\sqrt{\langle V\sb{ns}^2\rangle}$.}
\end{figure*}
We therefore reach an important conclusion, formulated in \Ref{PRB},  that the traditional form $\C P_1$ of the production term,  given by \Eq{prod-1}, contradicts both  the results of the numerical simulations in the channel and to the symmetry arguments.

To find a better  closure approximation  for the production term, let us replace  $\B V\sb{ns}\to  -\B V\sb{ns}$. It is natural to expect  that  $\<(\B s'\times\B s'')_x\>_{x,z}\to \<(\B s'\times\B s'')_x\>_{x,z}$, i.e. $\<(\B s'\times\B s'')_x\>_{x,z}$ is an even function of $\B V\sb{ns}$ (equal to zero for $\B V\sb{ns}=0$). Assuming analyticity, we conclude that for small $V\sb{ns}$ the function    $\<(\B s'\times\B s'')_x\>_{x,z}\propto V\sb{ns}^2$. If so, simple dimensional reasoning gives
\begin{equation}\label{est}
\<(\B s'\times\B s'')_x\>_{x,z}\ \sim  V\sb{ns}^2/ (\kappa ^2 \sqrt {\C L})\ .
\end{equation}
 We admit that these arguments are not rigorous. Nevertheless their conclusion\,\eqref{est} is  well  supported by the numerical simulation: See the solid and dashed blue lines in \Fig{f:2}. With the  closure\,\eqref{est}, \Eq{prod2} results in the form
 \begin{equation}\label{prod-2}
 \C P \Rightarrow \C P_3= \alpha C\sb{prod} \sqrt  {\C L} V\sb{ns}^3/\kappa^2\,,
 \end{equation}\end{subequations}
which was suggested in \Ref{PRB}.\\

Another comment of SN pertains to the decay term, which in his words ``can be extracted from dimensional analysis, which gives
\begin{subequations} \label{dec}
 \begin{equation}\label{decA}
 \C D=C \kappa \C L^2\ ."
  \end{equation}
In fact this is {\em not} the case. As was explained after Eq.(1) of \Ref{PRB}, this form was obtained by Vinen under the assumption that $\C D$ is independent of $V\sb{ns}$. The validity of this assumption is not obvious; the fact that this form is
applicable even for $V\sb{ns}\ne 0$, (as  demonstrated, e.g., in Fig.2b of \cite{PRB}),  is a nontrivial statement.  To clarify this point further, consider an analytical expression for $\C D$ that follows from \Eq{gen} in the Local Induction Approximation\,\cite{5,PRB}:
\begin{equation} \label{decB}
 \C D (y,t) = \frac{\alpha\kappa\, \ln(R/a_0)}{4\pi}\,\< | \B s ''|^2\>_{x,z}\ ,
 \end{equation}
 where $R$ is the mean radius of curvature.
 \end{subequations}
 This object  is again an even function of $V\sb{ns}$, but, in contrast to $\<(\B s'\times\B s'')_x\>_{x,z}$ it has a nonzero value for $V\sb{ns}=0$. Our support for the form\,\eqref{decA} of the decay term simply means that for the parameters of the numerical simulation in \Ref{PRB} the contribution proportional to $V\sb{ns}^2$ is small with respect  to the zero-order term\,\eqref{decA}.\\

In his Comment SN also expressed doubts about the flux term. In particular, citing \Ref{SN}, ``One more serious objection concerns the choice of flux  in the form Eq.(10) of the discussed paper. It is not motivated and looks strange."
To clarify the issue let us take  one step backward and recall that we possess a microscopic equation that describes the dynamics of the vortex tangle which as stated is not very useful for the coarse-grained macroscopical description. The goal is therefore to obtain a description that employs only a few macroscopic fields: the vortex-line density, the superfluid and normal fluid velocity. Obviously, this goal embodies a strong assumption; there is no guarantee that this is possible in every particular situation.

Consider a form for the flux term of the form of Eq.~\eqref{flux} below.  First, we stress that this form agrees perfectly with the  numerical results without any fitting parameters, see \Fig{f:4}. Second, we can explain
the origin of \Eq{flux}, by considering Eq.\,(6c) in our \Ref{PRB} for the flux (in the wall-normal direction $\B {\hat y}$),
\begin{subequations}\label{flux1}
\begin{equation}\label{flux1a}
\C J_y(y)= \C L(y)\< V\sp{drift}_y \>_{x,z}\ .
\end{equation}
Here the drift velocity is given by:
\begin{equation}\label{flux1b}
\bm V\sp{drift}=\bm V^s+(\alpha-\alpha'\bm s' ) \times \bm s' \times \bm V\sb{ns}
\end{equation}
The main contribution to the $y$ component of the drift velocity is the second term, caused by mutual friction, as is clearly seen in Fig. \ref{f:3}. In its turn, the main contribution to the friction term comes from the part proportional to $\alpha$. All this allows us to approximate the flux term as written in Eq.\,(6c) of \Ref{PRB},
\begin{equation}\label{flux1c}
\C J_y(y)\approx \alpha \C L(y)\< V_{{\rm ns},x} s'_z\>_{x,z}\approx \alpha V\sb{ns}\C L(y)\<  s'_z\>_{x,z}\ .
\end{equation}
In the last equation we took the slowly varying function $V_{ns}$ out of the average.

Next,  we note that  the only source of vorticity in superfluids is the quantized vortex lines. Therefore, $\kappa \C L(y) \<  s'_z\>_{x,z}$  is a $\B {\hat z}$ component of vorticity, calculated microscopically. On the other hand, on the macroscopic level, the vorticity is defined by $\B \nabla \times \B V\sb s$. By equating the $z$-components of these expressions, i.e.
 \begin{equation}\label{flux1d}
\kappa \C L(y) \<  s'_z\>_{x,z}= \frac{dV_{s}}{dy}\,,
\end{equation}\end{subequations}
we end up with an expression for the flux without any fitting parameters:
\begin{equation}\label{flux}
\C J_{y}(y,t)\approx  \frac{\alpha}{\kappa} V_{ns} \frac{dV_{s}}{dy}  \ .
\end{equation}

Finally, in the Introduction of \Ref{SN} SN  listed some steps in the analysis that he deems ``questionable."  These require further clarification.
\begin{enumerate}

\item SN noticed that
 ``The authors of \Ref{PRB}   have stated in the abstract of
their paper as ``To overcome this difficulty we announce here an approach that employs an inhomogeneous channel flow which is excellently suitable to distinguish the implications of the various possible forms of the desired
equation." SN refers to the difficulty of inferring dynamical equations from stationary flows.
Of course, in general, if one knows nothing about the system, it is difficult to say much about the temporal dependence studying only stationary states. But in our case we do know the microscopic equations that describe the dynamics of the vortex tangle. The whole problem discussed in \Ref{PRB} is how to estimate these expressions in terms of macroscopic variables.

\item In his Comment\,\cite{SN}  SN discussed  the limitations required to present an equation for $\partial \C L(\B r,t)$  in the closed form
\begin{equation}\label{2}
\frac{\partial \C L(\B r,t)}{\partial t} = \C F (\C L, V\sb{ns})\ .
\end{equation}
Clearly, in order to describe some averaged quantities defined by the configurations of the vortex tangle in terms of $V\sb{ns}$ and $\C L$, the statistics of the vortex tangle should be in a quasi equilibrium. Whether the same equation also describes
the equilibration process  is an open question that requires further study. In particular, in
  \Ref{PRB} we demonstrated that under the stated conditions the suggested  closure  form\,\eqref{2} exists and agrees
  very well with the numerics.

\item SN makes the statement that
the relaxation time of $\C L(t)$ is much larger than
that of $\B s''$; the latter has enough time to
adjust to the change in $ \C L(t)$, i.e.,  $|s''|\propto \sqrt{\C L}$.   Then, the self-preservation assumption is valid, and therefore the production term has
a classical form $\C P\propto |V\sb{ns}| \C L^{3/2}$.

We definitely disagree with this statement. The fact that $s''$ changes faster than $\C L$ does not imply that the production term has the classical form chosen by Vinen. This was explicitly shown in our \Ref{PRB}. The deep reason is that  $\C P$ is not proportional to $\<|s''|\>$. In reality $\C P\propto \< s'\times s''\>$. This term has completely different properties from  $\<|s''|\>$ as we explained above.

\end{enumerate}

To summarize, we have shown that reasonable  closure  estimates of the microscopic terms appearing in the equations
of motion of the density of vortex lines can be achieved. The resulting macroscopic equations were shown
to be in satisfactory agreement with detailed numerical simulations. We do expect that additional order parameters
may be required to describe more complex superfluid flows, but this is a challenge for the future.

\end{document}